\documentclass[showpacs,preprintnumbers,amsmath,amssymb,superscriptaddress]{revtex4}

\usepackage{graphicx}
\usepackage{dcolumn}
\usepackage{bm}

\begin{document}

\preprint{Physica A \textbf{388}, 977 (2009).}

\title{Stochastic resonance in time-delayed bistable systems driven by weak periodic signal}

\author{Rui-Hua Shao}
\affiliation{Institute of Theoretical Physics, Lanzhou University, Lanzhou $730000$, China}

\author{Yong Chen}
\altaffiliation{Corresponding author. Email: ychen@lzu.edu.cn}
\affiliation{Institute of Theoretical Physics, Lanzhou University, Lanzhou $730000$, China}

\date{\today}

\begin{abstract}
We study theoretically a bistable systems with time-delayed feedback driven by weak periodic force. The effective potential function and the steady-state probability density are derived. The delay time and the strength of its feedback can change the shapes of the potential wells. In the adiabatic approximation, the signal-to-noise ratio (SNR) of the system with a weak periodic force is obtained. The time-delayed feedback modulates the magnitude of SNR by changing the shape of the potential and the effective strength of signal. The maximum of SNR decreases with increasing the feedback intensity $\epsilon$. When $\epsilon$ is negative (or positive), the time delay can suppress (or promote) the stochastic resonance phenomenon.
\end{abstract}


\maketitle

In recent years, many complex systems with time-delay feedback have been investigated in optic laser systems, neural networks, coupled oscillators~\cite{delay1}, biological control, economic market, etc. In many cases, the delay reflects the transmission times related to the transport of matter, energy, and information through the system under investigation. Therefore, time-delay systems can be regarded as simplified but very useful descriptions of systems involving a reaction chain or a transport process. Noise and delayed time are two important elements in many complex systems. Such systems can be viewed as a special case of stochastic systems with memory, which have recently studied numerically~\cite{delay1,delay2} and analytically~\cite{delay3,delay4,delay5,delay6}.

Stochastic resonance (SR) behavior is one of the most studied and utilized fundamental physical phenomena and has already observed through experimental~\cite{SR2}. The original work on SR is mentioned by Benzi et al.~\cite{SR1}, for explaining the periodic recurrences of the earth's ice ages. The vast majority of studies on SR focus on the two-state model for symmetric systems~\cite{SR3,SR4,SR5,SR6} in the linear response theory (LRT). For a normal bistable system, the dynamics can be divided into two different regimes, the linear response regime $(A/D \ll1 )$ and the nonlinear regime $(A/D \gg 1)$ (where $A$ and $D$ are the amplitudes of the external periodic field and the noise strength, respectively). This point was discussed in~\cite{SR3,LRT,pbak5}. Shneidman et al.~\cite{wn1} studied the weak noise limit of SR in a bistable system. The SR of asymmetry systems have been also investigated widely by the authors of res.~\cite{pbak1,pbak2,pbak3,pbak4}. Later, Nikitin et al.~\cite{pbak5} expanded a comprehensive theory to study the effect of asymmetry on the switching dynamics in a bistable system with the limit of nonlinear response (weak noise limit). 

Recently, the time-delayed bistable system with small noise and magnitude of the feedback, has been widely studied in theory and experiment, especially the investigations on the residence time distribution and power spectra, having a great progress~\cite{delay7,delay8,delay9}. In the case of large delayed time ($\tau\approx 250$), Tsimring and Pikovsky calculated the correlation function and the power spectrum to predict the coherence resonance (CR) and SR behaviors~\cite{model1}. The phenomenon of CR was verified by experimentation~\cite{model2}. In addition, the studies on delayed noise (noise recycling) have also attracted more and more attention and already had great results~\cite{dn1,dn2}. 

In this paper, we study the model with small delay and weak periodic signal. This stochastic system is a delay-induced non-Markovian process. So it is hard to obtain a appropriate analytical result. Using the small-delay approximation of the probability density~\cite{delay4,delay5,pab1}, we derive the time-delayed Fokker-Planck equation and the effective Langvein equation. It is found that the SR indeed could be effected by delays from the analytical results for the effective potential function, the steady-state probability density, and the signal-to-noise ratio.

A typical time-delayed bistable systems is~\cite{model1}
\begin{equation}
\frac{dx(t)}{dt} = {x(t)} - {x^{3}(t)} + \epsilon{x(t-\tau)} + \sqrt{2D}{\xi(t)}. \label{eq-1}
\end{equation}
Here $\tau$ is the delay and $\epsilon$ is the strength of the feedback, which also can be regarded as a modulation parameter of the system itself. $\xi(t)$ is the Gaussian white noise with $\left< \xi(t) \right> = 0$ and $\left< \xi(t)\xi(t') \right> = \delta(t-t')$. $D$ is the strength of the noise.

We consider a classic model driven by a weak periodic signal with
the small delay
\begin{equation}
\frac{dx(t)}{dt} = {x(t)}-{x^{3}(t)}+\epsilon{x(t-\tau)}+A\cos(\Omega t) + \sqrt{2D}
{\xi(t)}, \label{eq-2}
\end{equation}
where $A$ is the strength of the signal (in general, we adopt $A=0.02$). For the low-frequency signal, the response of the systems to the weak periodic force is approximately linear, and the adiabatic limit is reasonable approach. The presence of small delay makes the bistable potential wells shack, but it is not enough to cause coherent resonance.

The dynamics in Eq.~(\ref{eq-2}) is a non-Markovian process. Using the probability density approach, the non-Markov process can be reduced to a Markov process and the approximate time-delayed Fokker-Planck equation is~\cite{delay4}
\begin{equation}
\frac{\partial P(x,t)}{\partial t} = -\frac{\partial [h_{eff} P(x,t)]}
{\partial x} + D\frac{\partial^{2}P(x,t)}{\partial^{2}x}.
\label{eq-3}
\end{equation}
Here the conditional average drift $h_{eff}$ reads
\begin{equation}
h_{eff} = \int^{b}_{a}dx_{\tau} h(x,x_{\tau})
P({x_{\tau},t-\tau}\mid{x,t}), \label{eq-4}
\end{equation}
where $x_{\tau}=x(t-\tau)$, $h(x,x_{\tau})=x-x^{3}+\epsilon x_{\tau}+
A\cos\Omega t$, $h(x)=x-x^{3}+\epsilon x+A\cos\Omega t$. The integral boundary $a$, $b$ tend infinite ($\pm\infty$). $P({x_{\tau},t-\tau}\mid{x,t})$ is the zeroth order approximate Markovian transition probability density~\cite{delay5,pab1}.
\begin{equation}
P({x_{\tau},t-\tau}\mid{x,t})=
\frac{1}{\sqrt{4\pi D\tau}} \exp \left( -\frac{(x_{\tau}-x-h(x)\tau)^{2}}{4D\tau} \right).
\label{eq-5}
\end{equation}
Substituting Eq.~(\ref{eq-5}) into Eq.~(\ref{eq-4}), we obtain
\begin{equation}
h_{eff}=(1+\epsilon\tau)(x-x^{3})+\epsilon(1+\epsilon\tau)x
+(1+\epsilon\tau)A\cos(\Omega t). \label{lab-6}
\end{equation}
So, the effective Langevin equation for Eq.~(\ref{eq-3}) becomes
\begin{eqnarray}
\frac{dx(t)}{dt}&=&(x-x^{3})+\epsilon x+A\cos(\Omega t)+\epsilon\tau [ (x-x^{3})\nonumber\\
&&+\epsilon x+A\cos(\Omega t) ] +\sqrt{2D}{\xi(t)}. \label{eq-7}
\end{eqnarray}

Note that a coupling term $\epsilon\tau \left[ (x-x^{3})+\epsilon x+A\cos(\Omega t) \right]$ is produced for the presence of time-delayed feedback. It indicates that the system is modulated by the delay and its feedback. From the effective Langevin equation obtained, one can investigate many dynamical properties of this system.

The unstable point $x_{0}$ and two stable points $x_{\pm}$ can be calculated easily. In the absence of the external periodic force, $x_{\pm}=\sqrt{1+\epsilon}$. It means that the delay cannot affect the positions of the unstable points and stable points.

The effective time-delayed potential function of Eq.~(\ref{eq-7}) is
\begin{equation}
U_{eff}(x) = -(1+\epsilon\tau)\left( \frac{1}{2}x^{2}-\frac{1}
{4}x^{4} \right)- \frac{1}{2}\epsilon(1+\epsilon\tau)x^{2} +(1+\epsilon\tau)A\cos(\Omega t)x, \label{eq-8}
\end{equation}
and the steady-state probability distribution function (PDF) is
\begin{equation}
P_{st}=N \exp \left( -\frac{U_{eff}}{D} \right),
\label{eq-9}
\end{equation}
where $N$ is the normalization constant, $N = \left(\int \exp (P_{st}) dx \right) ^{-1}$.

\begin{figure}
\includegraphics[width=0.4\textwidth]{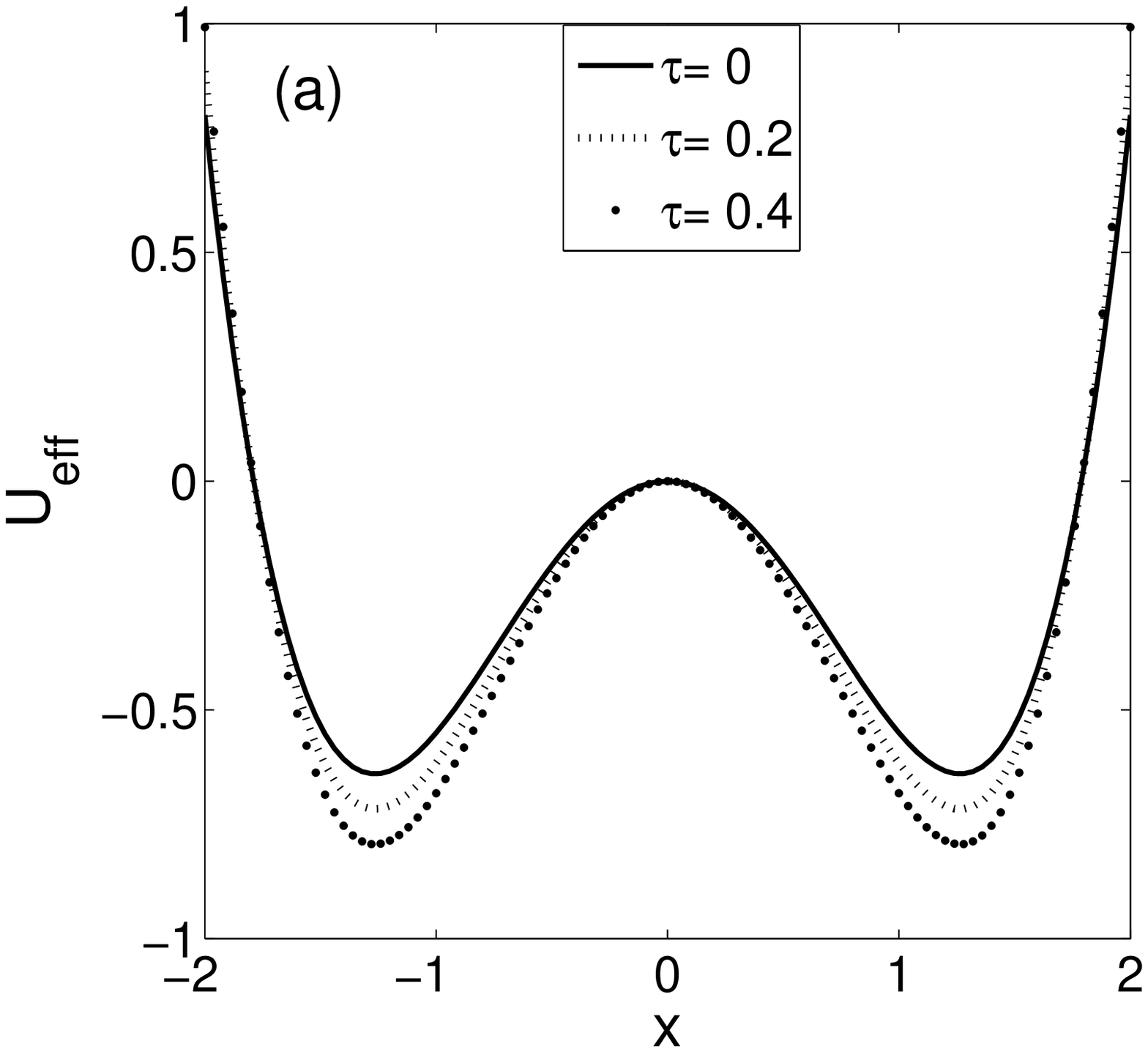}
\includegraphics[width=0.4\textwidth]{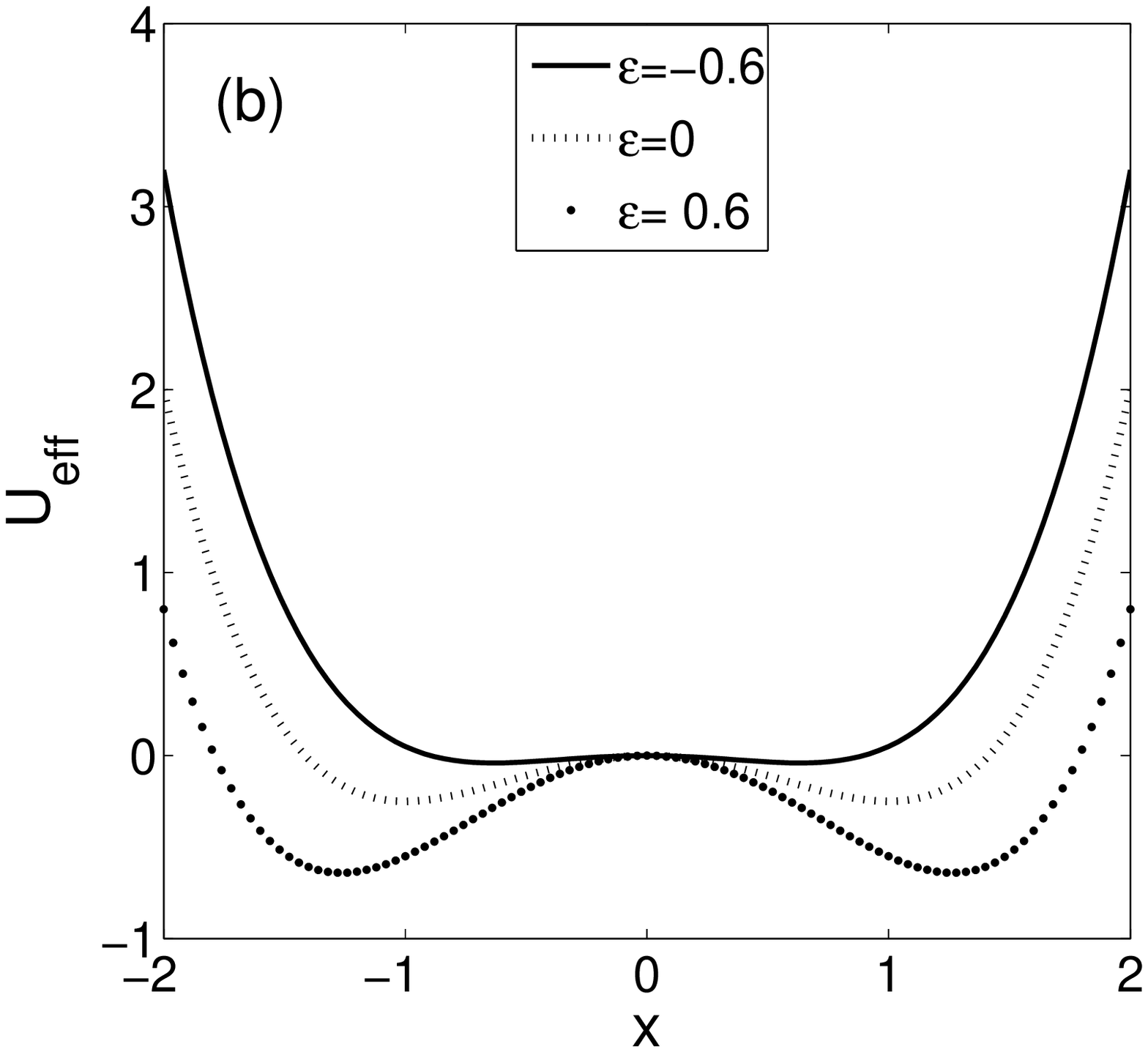}
\caption{(a) Potential $U_{eff}(x)$ for different delayed times with $\epsilon=0.6$. (b) Potential $U_{eff}(x)$ for different strengths of time-delay feedback with $\tau=0$. All the plots remove the effect of the external periodic forces.}
\label{fig-1}
\end{figure}

\begin{figure}
\includegraphics[width=0.4\textwidth]{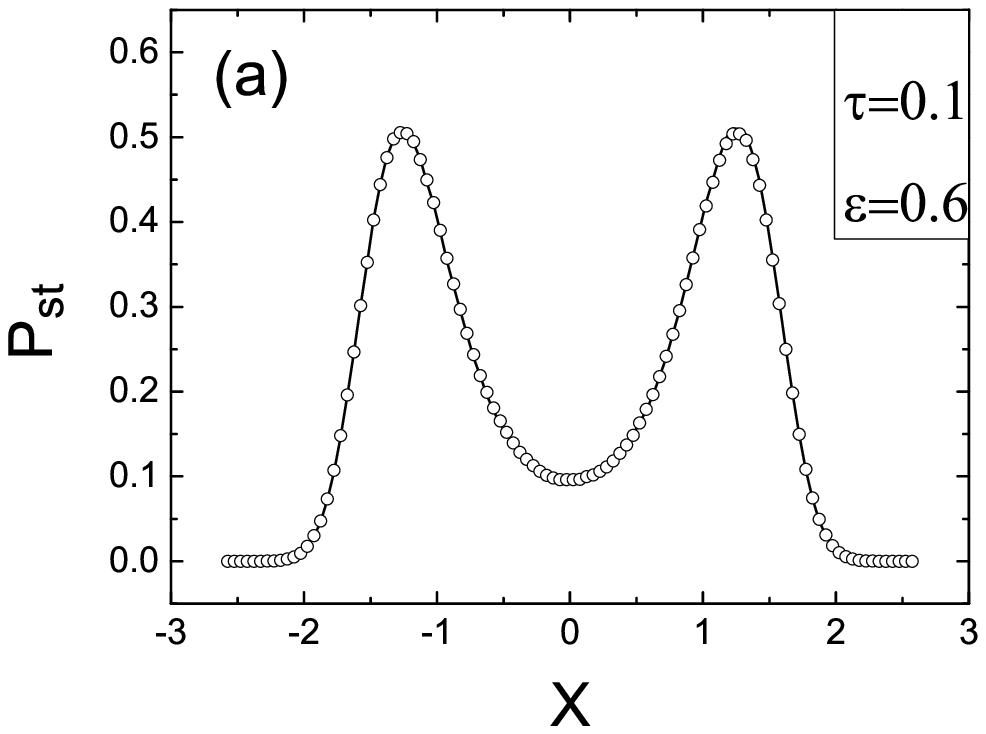}
\includegraphics[width=0.4\textwidth]{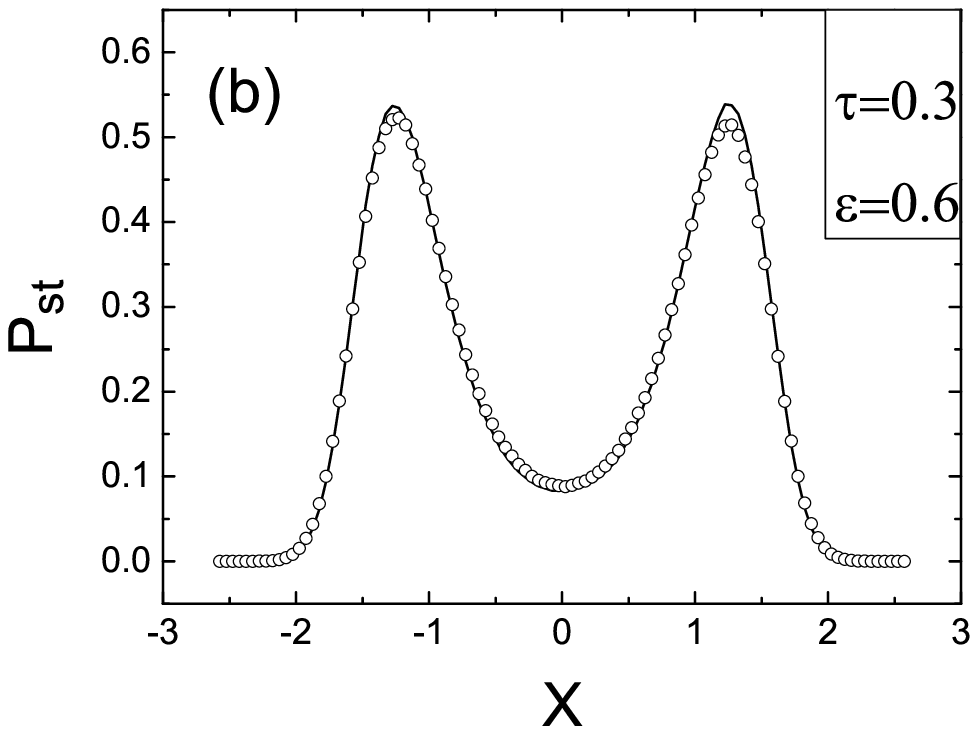}
\includegraphics[width=0.4\textwidth]{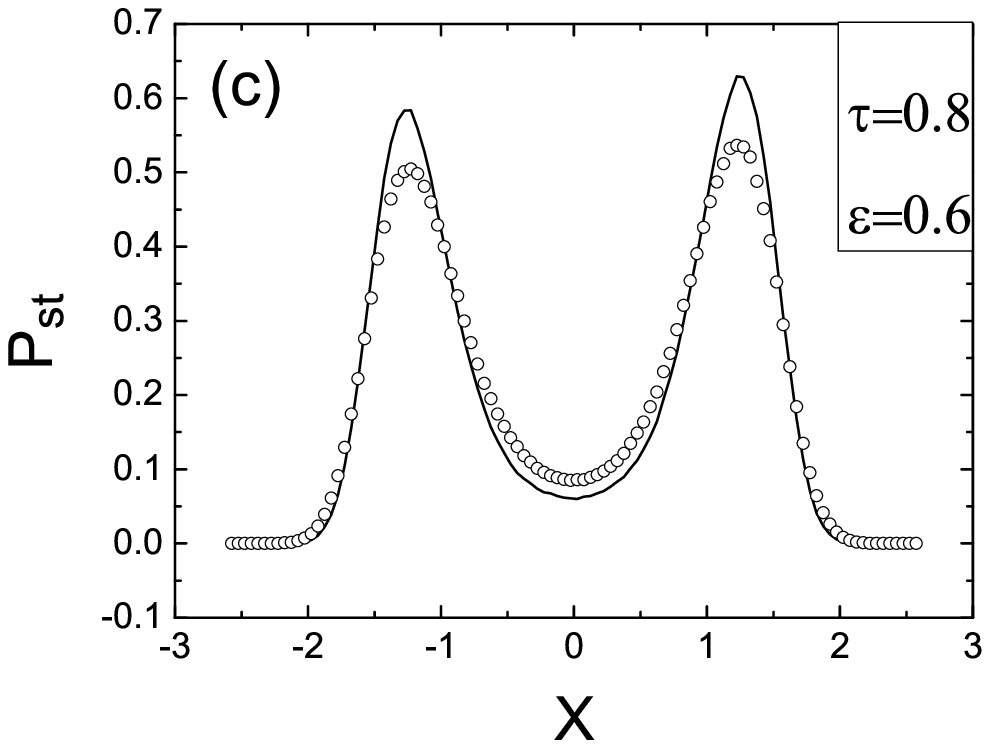}
\includegraphics[width=0.4\textwidth]{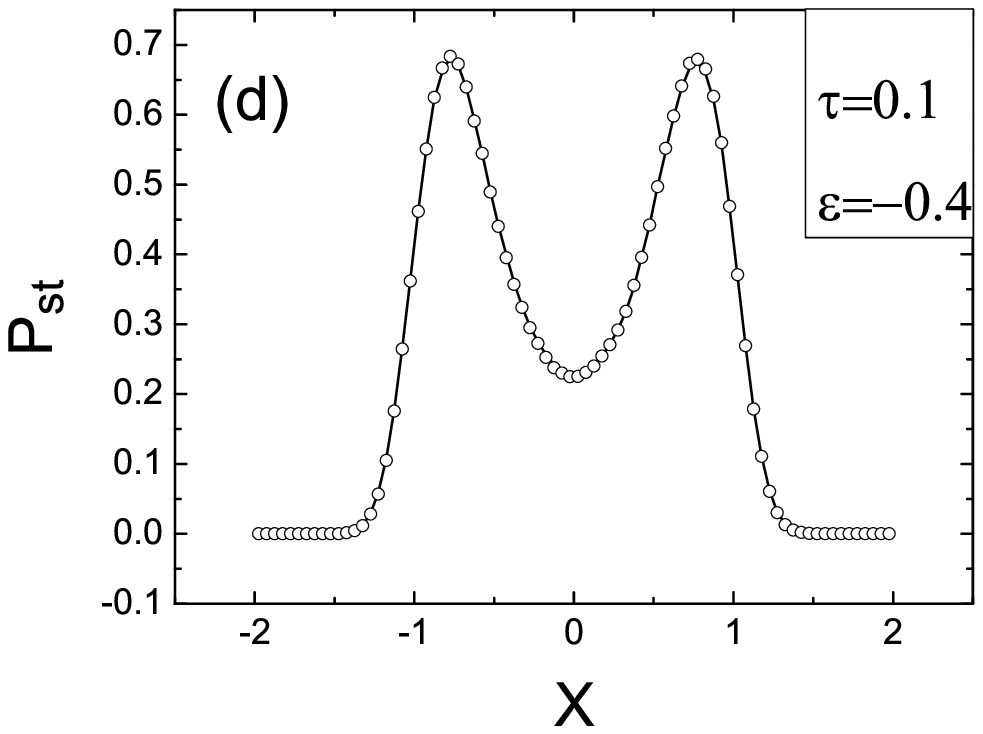}
\includegraphics[width=0.4\textwidth]{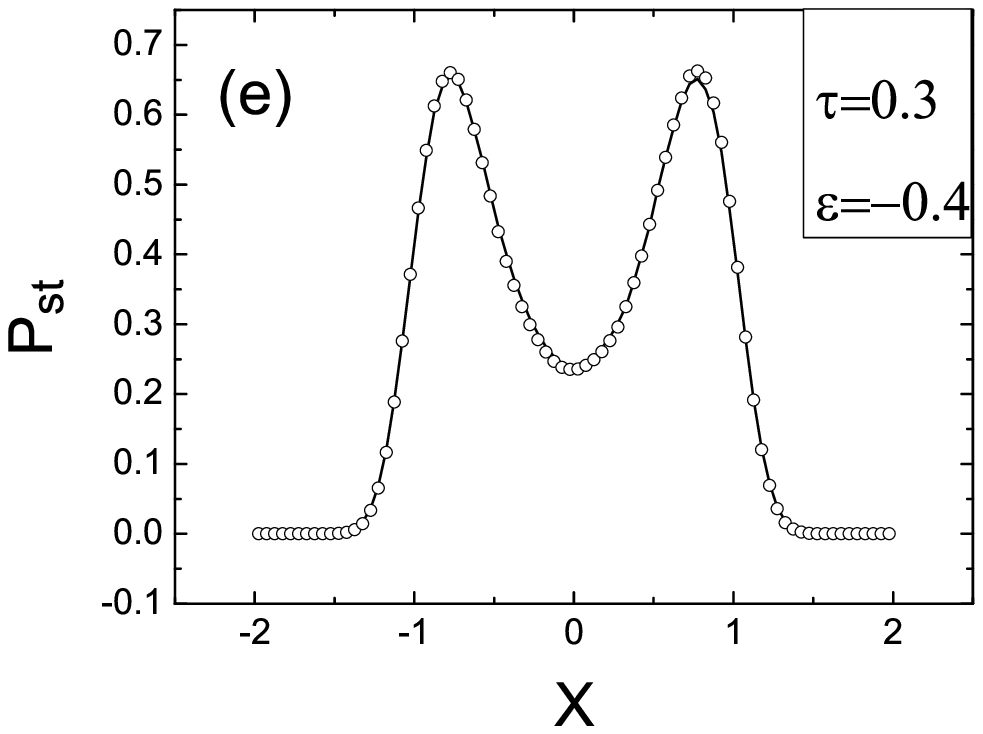}
\includegraphics[width=0.4\textwidth]{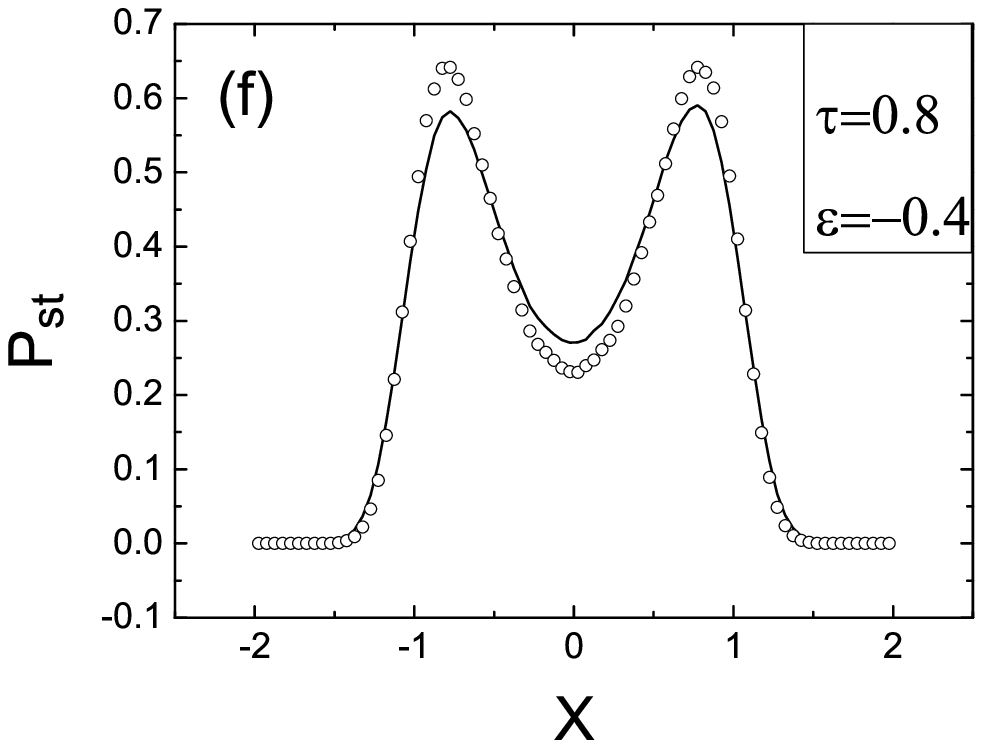}
\caption{The steady-state probability density $P_{st}$ for different delays $\tau$ with parameters $D$ and $\epsilon$: (a), (b), (c) $\epsilon=0.6$, $D=0.4$; (d), (e), (f) $\epsilon=-0.4$, $D=0.08$. The cycles represent the results obtained from simulations Eq.~(\ref{eq-2}), and the solid lines results form Eq.~(\ref{eq-7}), $A$=0.02.}
\label{fig-2}
\end{figure}

\begin{figure}
\includegraphics[width=0.5\textwidth]{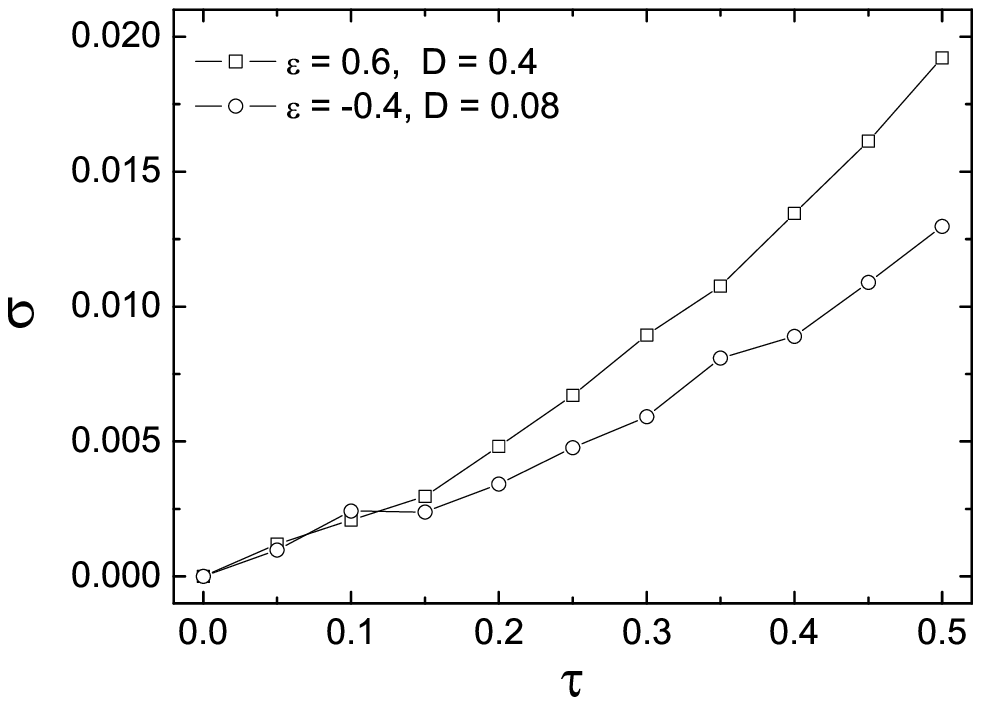}
\caption{The standard deviation (SD) $\sigma=\sqrt{1/(n-1)\sum_{i=1}^n(p_{ia}-\overline{p}_{is})^{2}}$. $p_{ia}$ is the analytical value of $P_{st}$ at the $i$th point, and $\overline{p}_{is}$ is the average value of $P_{st}$ from simulation at $i$th point.}
\label{fig-3}
\end{figure}

In this model, the modulation parameter $\epsilon$ is expected to be larger than $-1$ (or should be larger) to ensure the bistability of the time-dependent potential $U(x,t,\tau)$ possessing minima, respectively, at $x_\pm$. Moreover, the depth and width of the potential wells increase with increasing of the strength of $\epsilon$, since the particles need larger noise $D$ to cross the barrier. In order to limit the size of $D$, $\epsilon$ is also not too large. In general, the positive $\epsilon$ is not more than $1$ in this paper. All viewpoints are discussed on this basis.

Without the periodic force, we depict the forms of potential for different time delays and their feedbacks in Fig.~\ref{fig-1}(a, b). It is found that the depth of the potential wells increases with increasing of time delays for $\epsilon>0$, but the width between two wells (two peaks) remains unchanged. That is to say the peak of the PDFs becomes sharper for deeper potential wells by increasing the ability of capturing particles. It means that the peak values of $P_{st}$ increase with increasing of time delays. However, for $\epsilon<0$, the results are opposite. In Fig.~\ref{fig-2}(a, b), when $\epsilon=0.6$, $\tau=0.1$ and $0.3$, the peak value $P_{max}$ from the analysis is ($0.508$, $0.539$), and from the simulation is ($0.505$, $0.523$). Correspondingly, for $\epsilon=-0.4$, $P_{max}$ is: from the analysis $(0.684$, $0.663)$, and from the simulation $(0.681$, $0.657)$. The results are consistent with the above discussion. Fig.~\ref{fig-1}(b) shows that the depth and width of the potential wells increase with increasing of the strength of time-delayed feedback $\epsilon$, therefore the particles need larger noise to cross the barrier (see Fig.~\ref{fig-2}).

Fig.~\ref{fig-2} and Fig.~\ref{fig-3} show the steady-state PDFs from the analysis and numerical simulations and their standard deviation (SD) $\sigma$. It is found that the small-delay approximation of the probability density is a good approximation in the case of small delay (corresponding to $\sigma<0.01$). For the smaller noise, the availability of the approach is much better. In Fig.~\ref{fig-2}, the steady-state probability density of Eq.~(\ref{eq-2}) (the circles) is consistent with Eq.~(\ref{eq-7}) (the solid line) for $\tau<0.4$, especially within $\tau=0.3$.

In general, the dynamics of a bistable system can be divided into two different regimes, the linear response regime and the the nonlinear regime. In the limit $A\ll D$, the response of the system to the periodic force is approximately linear (regarding $A$ as perturbation term) and the linear response theory can be applied. In our systems, both the transition time of the probability from one well to another one and the signal period are much longer than the relaxation time of the system. As a result, the adiabatic limit works valid too.

From Eq.~(\ref{eq-7}), note that the unstable and stable points of the system are not influenced by the time delay $\tau$. It indicates that the particle spends most of the time near the minimal potential $x = \pm \sqrt{1+\epsilon}$, occasionally jumping from one to another because of the noise. The transition rates between the wells are Kramers' escape rates~\cite{SR3,Kram}
\begin{equation}
\gamma_{\pm} = \frac{(1+\epsilon)(1+\epsilon\tau)} {\sqrt{2}\pi} \exp \left[ -\frac{(1+\epsilon)^{2}(1+\epsilon\tau)}{4D} \right] \times \exp \left[
\frac{\pm(1+\epsilon\tau)\sqrt{1+\epsilon}A\cos \Omega t}{D}
\right]. \label{eq-10}
\end{equation}
Here, $\gamma_{+}$ is the transition rate of the probability from left-hand well to the right-hand, and vice versa.

Normally, the periodic force is small $(Ax \ll \bigtriangleup U)$ and slow $(\Omega\ll\omega_{0})$, where $\omega_{0}$ is the frequency of system vibrations inside the well. In this paper, the frequency of signal $\Omega=0.003$. Without the periodic force, we have the approximation of Eq.~(\ref{eq-10}),
\begin{equation}
\gamma_{\pm 0} = \frac{(1+\epsilon)(1+\epsilon\tau)}{\sqrt{2}\pi} \exp \left[-\frac{(1+\epsilon)^{2}(1+\epsilon\tau)}{4D} \right].
\label{eq-11}
\end{equation}
Consequently, the Kramers' escape time is $\gamma_{0}^{-1}$. In general, the time delay $\tau\ll\gamma_{0}^{-1}$.

The power spectrum density of the output variables, a Fourier transform of the autocorrelation function, in Eq.~(\ref{eq-7}) is given by~\cite{SR3,SR4,SR5}
\begin{equation}
S(\omega) = \frac{\pi (1+\epsilon) M^{2}} {2(N^{2}+\Omega^{2})}
\left[ \delta(\Omega-\omega) + \delta(\Omega+\omega) \right] + \left[ 1-\frac{M^{2}}{2(N^{2}+\Omega^{2})} \right]
\frac{2(1+\epsilon)N}{N^{2}+\omega^{2}}, \label{eq-12}
\end{equation}
where
\begin{eqnarray}
N&=&\gamma_{+0}+\gamma_{-0} = \frac{\sqrt{2}(1+\epsilon)(1+\epsilon\tau)}{\pi} \exp \left
( -\frac{(1+\epsilon)^{2}(1+\epsilon\tau)}{4D} \right),\nonumber\\
M&=&\frac{A}{D}(1+\epsilon\tau)\sqrt{1+\epsilon}N.
\label{eq-13}
\end{eqnarray}
In Eq.~(\ref{eq-12}), $A\ll1$ allow us to omit the last term $-M^{2}/\left[ {2(N^{2}+\Omega^{2}}) \right]$. So $S(\omega)$ is defined only for positive $\Omega$. Then, Eq.~(\ref{eq-12}) becomes
\begin{eqnarray}
S(\omega) &=& S_{1}(\omega) + S_{2}(\omega), \nonumber \\
S_{1}(\omega) &=& \frac{\pi(1+\epsilon)M^{2}} {2(N^{2}+\Omega^{2})}
\delta(\Omega-\omega), \nonumber\\
S_{2}(\omega) &=& \frac{2(1+\epsilon)N}{N^{2}+ \omega^{2}},
\label{eq-14}
\end{eqnarray}
where $S_{1}(\omega)$ is the output power spectrum of the signal,
and $S_{2}(\omega)$ is of noise. From the definition of SNR,
$R=S_{1} (\omega)/S_{2}(\omega)$, we have
\begin{equation}
R = \frac{\sqrt{2}A^{2}}{4D^{2}}(1+\epsilon)^{2}(1+\epsilon\tau)^{3}
\exp \left[ -\frac{(1+\epsilon)^{2}(1+\epsilon\tau)}{4D}
\right].
\label{eq-15}
\end{equation}
\begin{figure}
\includegraphics[width=0.4\textwidth]{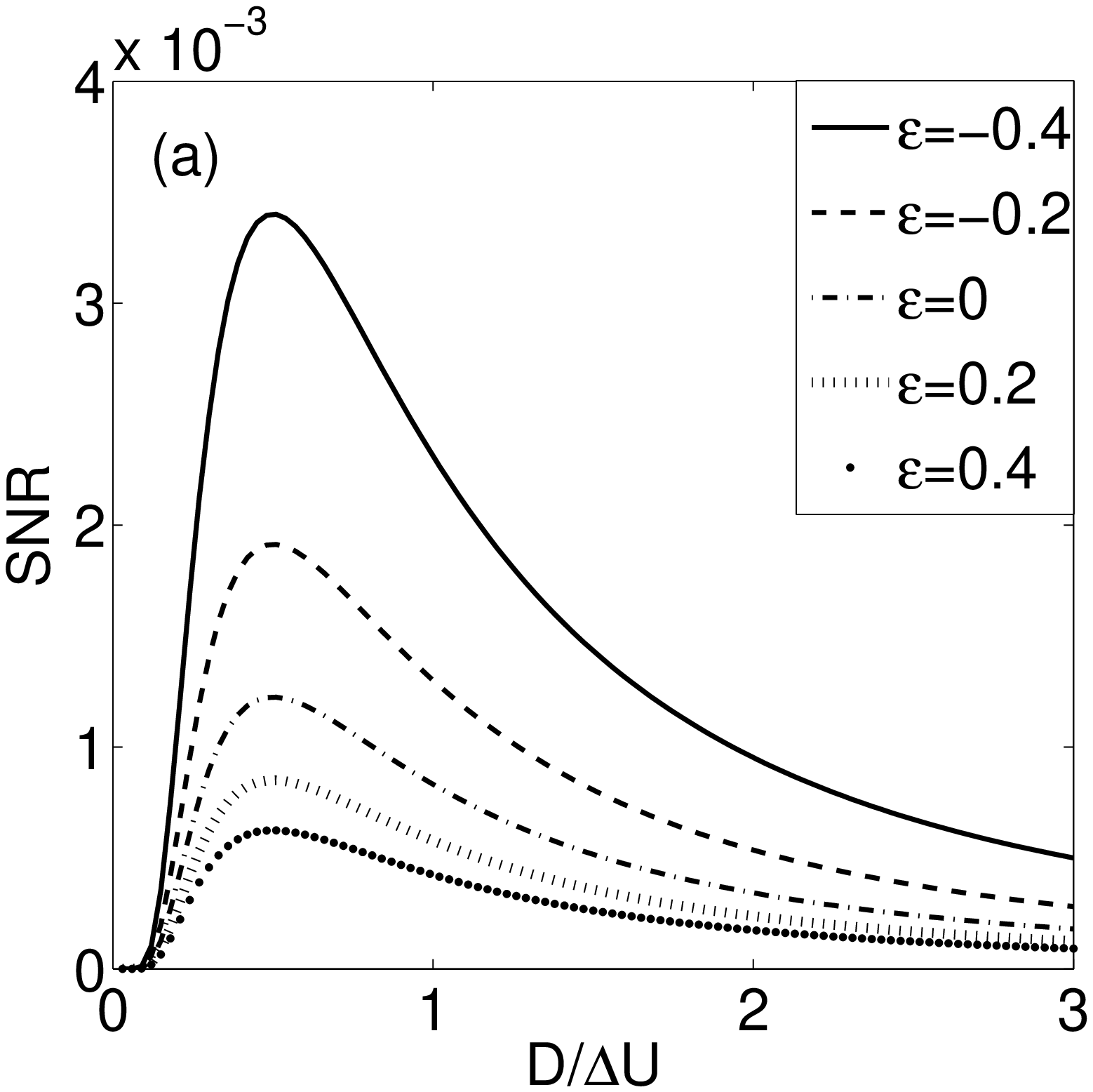}
\includegraphics[width=0.4\textwidth]{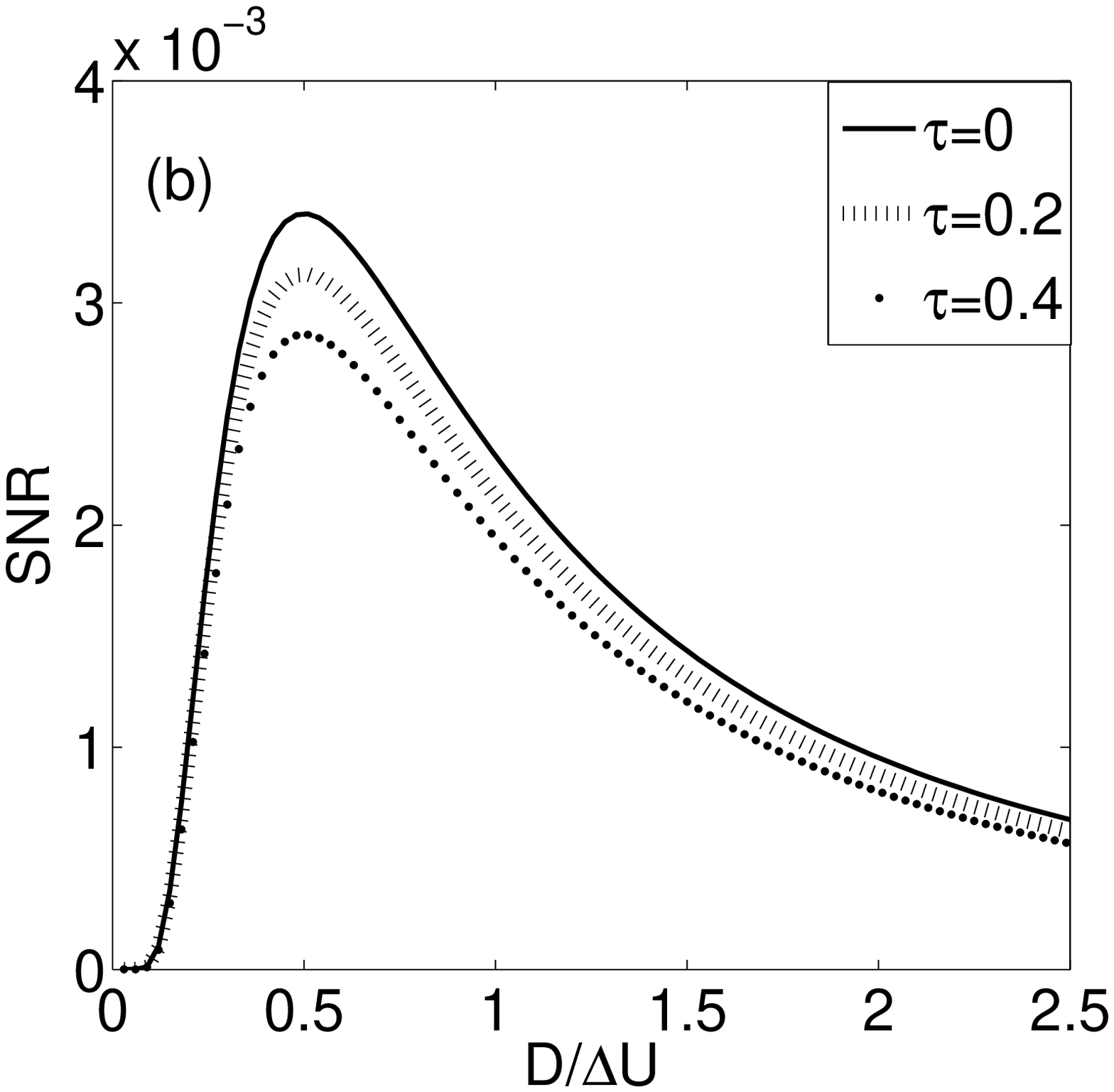}
\includegraphics[width=0.4\textwidth]{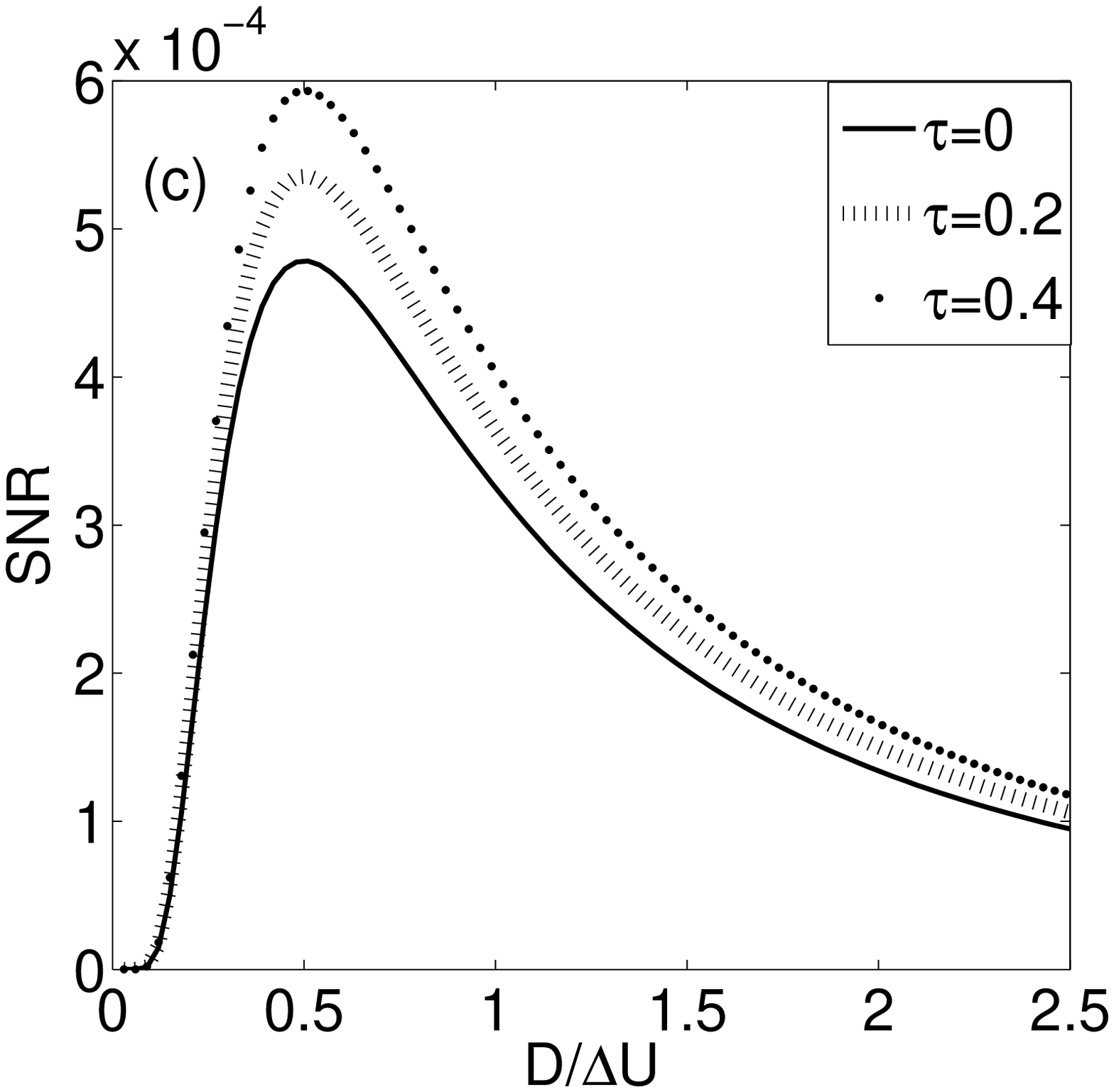}
\caption{SNR as a function of the noise intensity $D/\Delta U$ with the magnitude of the signal $A=0.02$, where $\Delta U$ is the height of the potential barrier without external periodic force. Parameters are: (a) $\tau=0$, (b) $\epsilon=-0.4$, (c) $\epsilon=0.6$.}
\label{fig-4}
\end{figure}

From $dR/dD = 0$, we get
\begin{equation}
D =\frac{1}{8}(1+\epsilon)^2 (1+\epsilon \tau) = \frac{1}{2}\Delta U. \label{eq-16}
\end{equation}
Here, $\Delta U$ is the height of potential barrier. When $D/\Delta U \approx 0.5$, the SNR reaches a maximum.

In Fig.~\ref{fig-4} we show the curves of SNR as a function of the noise intensity $D$ for different $\epsilon$ and $\tau$. Note that each curve exhibits an optimum noise intensity where the SNR has a maximum, which is the characteristic signature of SR phenomenon. Furthermore, it is obvious that the peak value of SNR decreases with increasing of feedback $\epsilon$ when $\tau$ is a constant [see Fig.~\ref{fig-4}(a)]. In the case of negative feedback, the time delay suppresses the SR phenomenon, on the contrary, the larger delay will help to enhance the peak of SNR for the positive feedback [as shown in Fig.~\ref{fig-4}(b)-(c)].

\begin{figure}
\includegraphics[width=0.4\textwidth]{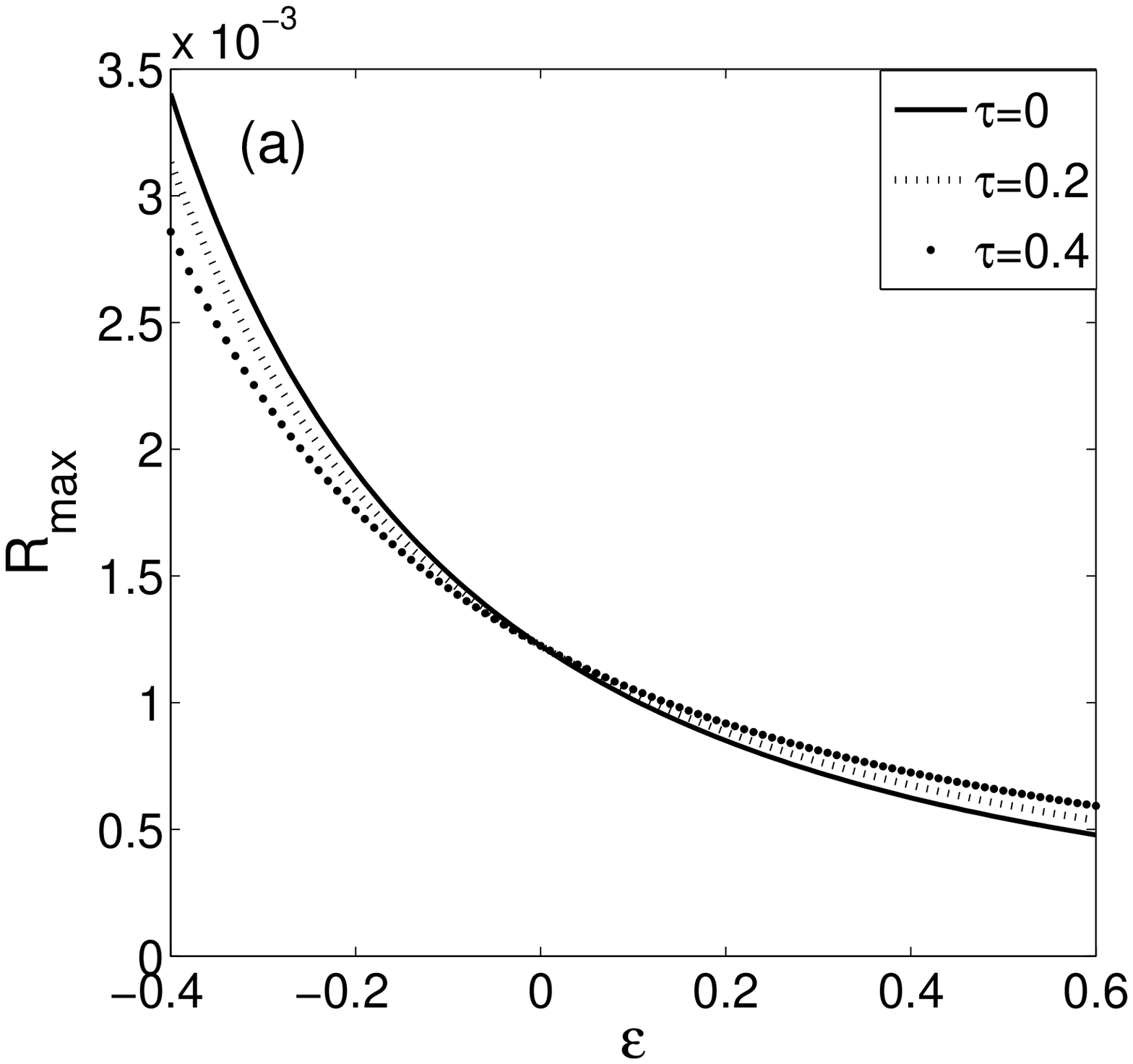}
\includegraphics[width=0.4\textwidth]{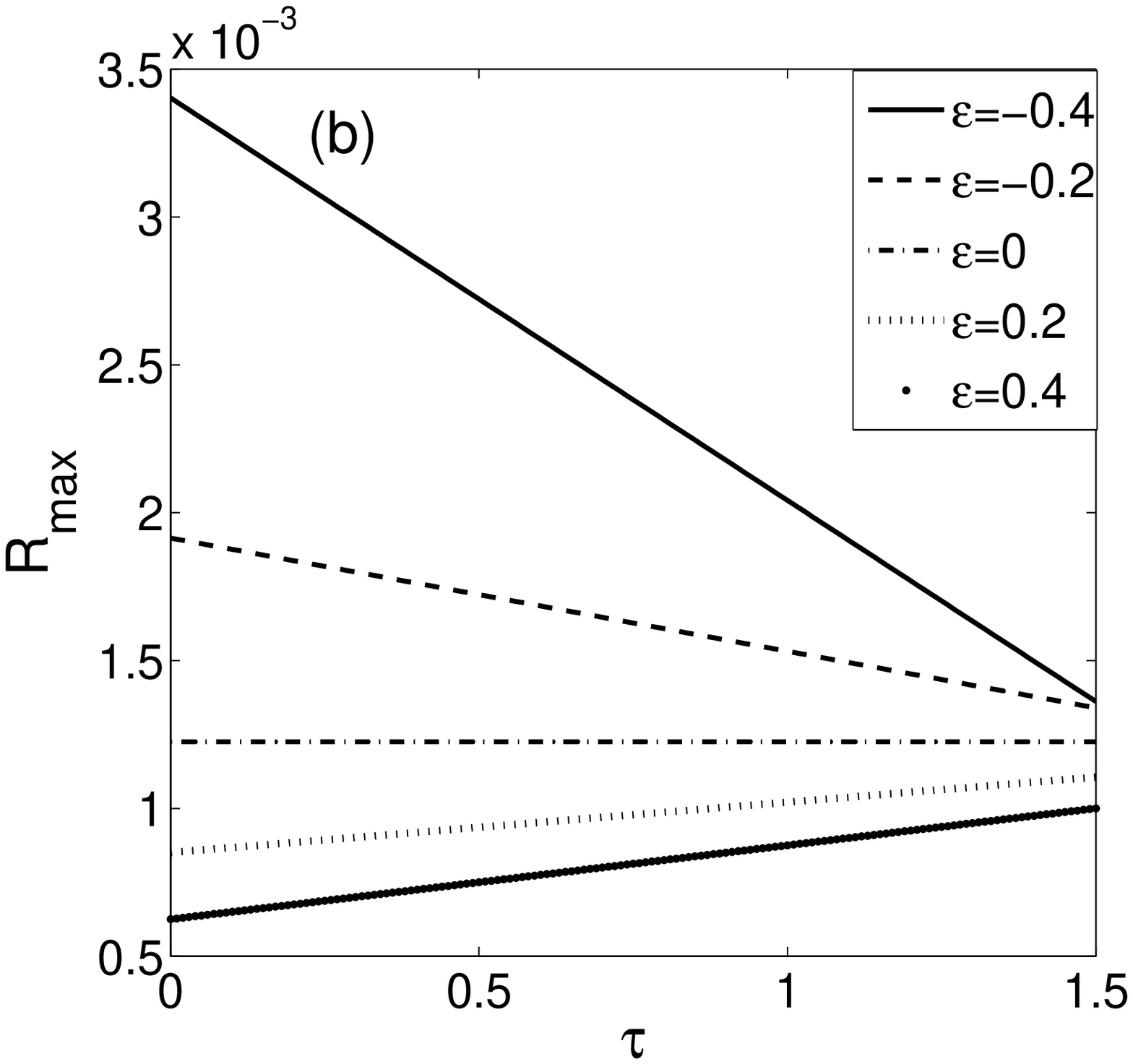}
\caption{(a) $R_{max}$ as the function of $\epsilon$; (b) $R_{max}$ as the function of $\tau$.} \label{fig-5}
\end{figure}

It is interesting to compare the effects of delay and its feedback on $R_{max}$ (the maximum of SNR ). When $D/\Delta U=0.5$, the maximum of SNR is
\begin{equation}
R_{max} = 16 \sqrt{2} e^{-2} A^2 \frac {1+\epsilon\tau}{(1+\epsilon)^{2}}. \label{eq-17}
\end{equation}
The model [Eq.~(\ref{eq-2})] is controlled within a small delay. Note that $\epsilon=0$ is the critical point. The negative $\epsilon$ is more effective to promote the SR than the positive [see Fig.~\ref{fig-5}(a)]. In Eq.~(\ref{eq-17}) and Fig.~\ref{fig-5}(b), the linear relationship of $R_{max}$ and $\tau$ is displayed when $\epsilon$ is fixed. Obviously, within a small delay, it is appropriate to observe the clear SR phenomenon when the feedback is negative.

Although the delay $\tau\ll\gamma^{-1}$ and $\tau\ll T$ (where $\gamma^{-1}$ and $T$ are Kramers' escape time and the period of the signal), the effect of time delay on SR cannot be neglected. Eq.~(\ref{eq-2}) can be rewritten by the following form
\begin{equation}
\frac{dx(t)}{dt} = (1+\epsilon){x(t)}-{x^{3}(t)}+\epsilon[{x(t-\tau)}-{x(t)}]+A\cos(\Omega t) + \sqrt{2D} {\xi(t)}, \label{eq-18}
\end{equation}
where $\epsilon[{x(t-\tau)}-{x(t)}]$ is a small quantity. We define
\begin{equation}
\epsilon[{x(t-\tau)}-{x(t)}]= \epsilon\chi(\tau)x(t), \label{eq-19}
\end{equation}
as the function of $\tau$. $\chi(\tau)$ is a small quantity too, and is a linear relationship with the delay. Then the another form of Eq.~(\ref{eq-2}) is presented,
\begin{equation}
\frac{dx(t)}{dt} = [1+\epsilon+\epsilon\chi(\tau)] x(t)-x^{3}(t)+A\cos(\Omega t) + \sqrt{2D} {\xi(t)}. \label{eq-20}
\end{equation}
After the moment of a particle crossing the barrier, from the left well to the right, ${x(t-\tau)}<{x(t)}$, $x(t)>0$, then $\chi(\tau)<0$; contrarily, when the particle just jump into the left well, $x(t)<0$, ${x(t-\tau)}-{x(t)}>0$, and $\chi(\tau)<0$.

It is well known that the depth of the potential wells $\Delta U$ (or the potential barrier) is one of the important parts to contribute to SNR. In our model, the amplitude of the signal $A$ is a constant, thus its ability to modulate wells will keep unchanged. When $\Delta U$ increases with $\epsilon$, the particles are difficult to cross the barrier, which prevents the noise and signal from achieving resonance. Besides, with the enhancement of the noise background $S_{2}(\omega)$ (for keeping $D/\Delta U \approx 0.5$), SNR will weaken. It means that the peak of SNR will decrease with the raising of $\epsilon$, which is shown in Fig.~\ref{fig-4}(a).

When the delay $\tau$ exists in this system, at the moment of a particle just crossing the barrier, the function $\chi(\tau)<0$. For the negative feedback $\epsilon$, the term $\epsilon\chi(\tau)>0$, $1+\epsilon+\epsilon\chi(\tau)>1+\epsilon$, which causes the potential barrier $\Delta U$ larger in Eq.~(\ref{eq-20}), thus the peak value of SNR will smaller [see Fig.~\ref{fig-4}(b)]. That is to say that the delay $\tau$ prevents the particle returning to the former well after it just reach one well, and this behaviors lead to SR phenomenon weakening. Whereas the term $\epsilon\chi(\tau)<0$ with the positive $\epsilon$, $1+\epsilon+\epsilon\chi(\tau)<1+\epsilon$, the potential barrier $\Delta U$ is depressed, and the particle can easily return to the original potential well. So $\tau$ is helpful to raise the peak of SNR [shown in Fig.~\ref{fig-4}(c)].

From the part induced by the time delay $\tau$ in Eq.~(\ref{eq-7}), $\epsilon\tau \left[ (x-x^{3})+\epsilon x+A\cos(\Omega t) \right]$, the external periodic signal and the original system are modulated together by the time delay $\tau$ and its feedback intensity $\epsilon$. Although $\tau$ can tune the potential barrier $\Delta U$ from the negative (or positive) feedback $\epsilon$, the amplitude of the signal $A^{*}$ is also modulated by $\epsilon$ and $\tau$, $A^{*}=(1+\epsilon\tau)A$. So the variations of SNR displayed by Fig.~\ref{fig-4} are proper.

Generally speaking, in the stochastic system driven by weak periodic force, because of the presence of the delay $\tau$ and the strength of its feedback $\epsilon$, the system produces a coupling term, such as $\epsilon\tau \left[(x-x^{3})+\epsilon x+A\cos(\Omega t) \right]$ in this work. By changing $\tau$ and $\epsilon$, the signal amplitude $A$ is modulated, and the peak value $R_{max}$ of SNR exhibits different magnitudes under the same intensity of noise ($D/\Delta U \approx 0.5$) and shows different tendency. When $\tau$ keeps unchanged, the SR phenomenon of the negative $\epsilon$ ($-1<\epsilon<0$) is more easily observed than it for the positive $\epsilon$. For the positive $\epsilon$, the SNR of the system is promoted with increasing of the time delay $\tau$. On the contrary, the SR phenomenon with the negative $\epsilon$ is suppressed by $\tau$.

In this paper we study a simple time-delayed bistable model with Gaussian white noise driven by weak periodic force. Using the small delay approximation of the probability density function, the time-delayed Langevin equation is extended to a effective Langevin equation. Correspondingly, we obtain the effective potential function $U_{eff}$ and the steady-state probability density $p_{st}$. For a weak periodic force, the analytical expressions of SNR and $R_{max}$ are derived in the adiabatic limit, and the influences of delays on them are also discussed. It shows that the time delay is one of the most important elements to affect the dynamics of complex systems.

\begin{acknowledgments}
We really appreciate the anonymous referees for their very constructive and helpful suggestion, and acknowledge simulation discussions with S. M. Qin, W. K. Qi, and L. C. Yu. This work was supported by the National Natural Science Foundation of China under Grant No. $10305005$ and by the Fundamental Research Fund for Physics and Mathematic of Lanzhou University.
\end{acknowledgments}

\end{document}